# Experimental Indications for the Response of the Spectators to the Participant Blast

M. V. Ricciardi[a], T. Enqvist[a,*], J. Pereira[b], J. Benlliure[b],

M. Bernas[c], E. Casarejos[b], V. Henzl[a,†], A. Kelić[a], J. Taïeb[c,‡], K.-H. Schmidt[a]

[a]*GSI, Planckstr. 1, 64291, Darmstadt, Germany*
[b]*University of Santiago de Compostela, 15706 Santiago de Compostela, Spain*
[c]*Institut de Physique Nucléaire, 91406 Orsay Cedex, France*

**Abstract:** Precise momentum distributions of identified projectile fragments, formed in the reactions $^{238}$U + Pb and $^{238}$U + Ti at 1 $A$ GeV, are measured with a high-resolution magnetic spectrometer. With increasing mass loss, the velocities first decrease as expected from previously established systematics, then level off, and finally increase again. Light fragments are on the average even faster than the projectiles. This finding is interpreted as the response of the spectators to the participant blast. The re-acceleration of projectile spectators is sensitive to the nuclear mean field and provides a new tool for investigating the equation of state of nuclear matter.

*I. Introduction.* -- The equation of state of nuclear matter, responsible for its response to temperature and density, belongs to the key topics of nuclear physics. Besides nuclear-physics aspects it touches on important questions in astrophysics and cosmology, e.g., the dynamics of supernova explosions [1], the stability of neutron stars under gravitational pressure [2] and the nature of matter that existed in the early universe [3]. Intensive effort has been invested to extract information on the equation of state and on in-medium nucleon-nucleon interactions by observing light particles emerging from central collisions in full-acceptance experiments (e.g. [4]). Also the statistical behaviour of the spectator matter has been intensively investigated [5, 6]. Recently, it has been brought into the discussion [7] that the kinematical properties of the spectators in mid-peripheral collisions could carry important complementary information to the results of the above-mentioned experiments. According to these model calculations [7], the transversal and the longitudinal momentum distributions of the spectators are influenced by the participant blast, occurring after the compression phase in the colliding zone, and thus the momentum distributions are sensitive to the nuclear force. However, the momentum distributions have to be measured with high precision in order to yield conclusive results.

A review on measured mean velocities of spectator-like fragments has been presented by Morrissey in 1989 [8]. This systematics shows a clear correlation of the observed momentum shift with the mass loss in very peripheral collisions which can be interpreted as the consequence of a kind of friction in the nucleus-nucleus collision. The momentum distributions of lighter fragments, with a mass loss larger than about 1/3 of the mass of the projectile, respectively the target nucleus, however, show a large spreading with no clear tendency. According to the model calculations of ref. [7], these are the products, which carry

---

the most valuable information on the basic nuclear properties mentioned above. Some previously measured data are especially interesting in this context. Using a thick-target thick-catcher technique, target-like fragments around A = 60, produced in the bombardment of gold by $^{12}$C at 25 GeV [9], were found to be produced with mean velocity close to zero in the laboratory frame. An experiment of Loveland et al. [10], based on the same technique, revealed a backward emission of $^{24}$Na, produced by multi-fragmentation of gold by $^{16}$O at 232 GeV. Results from a time-of-flight measurement [11] in a comprehensive projectile-fragmentation experiment with a $^{197}$Au beam of 600 $A$ MeV and targets from carbon to lead showed a levelling off of the velocities of the fragments with increasing mass loss. Fragments below $Z$ = 50 even showed a tendency to come closer to the beam velocity again.

In this situation, we made use of a high-resolution magnetic spectrometer to obtain experimental data on momentum distributions of fully identified reaction residues with high precision. The magnetic rigidity of each residue can be measured with high precision just by determining the deflection in a magnetic dipole field. This gives very precise information on its longitudinal momentum and hence on its velocity, once the product is identified in mass and atomic number.

*II. Experiment.* -- The SIS18 heavy-ion accelerator of GSI, Darmstadt, was used to provide a $^{238}$U beam of 1 $A$ GeV with an average intensity of about $10^7$ particles per second. The beam impinged on two different targets, a 36 mg/cm$^2$ titanium target and a 50.5 mg/cm$^2$ lead target. In the target layers used in the two experiments, the primary beam looses less than 1% of its energy; thus corrections due to energy loss in the target do not deteriorate the accurate measurement of the longitudinal momenta of the reaction residues. The reaction products entered into the fragment separator [12], used as a high-resolution magnetic spectrometer. Full identification in mass and atomic number of the reaction residues was performed by determining the mass-over-charge ratio $A/Z$ from magnetic rigidity and time-of-flight, and by deducing the atomic number $Z$ from a $\Delta E$ measurement with an ionisation chamber. Once the reaction residue was identified, the measurement of the magnetic rigidity, deduced from the horizontal position at the intermediate dispersive image plane of the fragment separator, gave precise information on its longitudinal momentum. The acceptance of the fragment separator is about 3 % in momentum and 15 mrad in angle around the beam axis. Measurements with different magnetic fields were combined to fully cover the momentum distributions of all residues. Details of the experimental set-up and the detector equipment [13, 14], as well as a description of the analysis method [15, 16, 17] are given in previous publications. The magnetic fields were measured by Hall probes with a relative precision of $10^{-4}$. The bending radius was deduced from the position at the intermediate image plane with a statistical relative uncertainty of $\pm 5 \cdot 10^{-4}$, based on a resolution of $\pm 3$ mm in the position measurement. This results in a relative uncertainty of $\pm 5 \cdot 10^{-4}$ in the momentum of individual reaction products.

*III. Results.* -- The essential result of the experiment is the systematic scan of the velocity distributions of all reaction residues in the region of interest. For the system $^{238}$U + Pb, all elements between vanadium ($Z$ = 23) and rhenium ($Z$ = 75) were measured, while for the system $^{238}$U + Ti an even larger range from oxygen ($Z$ = 8) to uranium ($Z$ = 92) was covered. Figure 1 shows two-dimensional cluster plots of the velocity distributions of the isotopes of eight elements from the reaction $^{238}$U + Pb, measured at 1 $A$ GeV. The picture must be interpreted keeping in mind the limited angular acceptance of the spectrometer. The angular transmission for projectile fragments ranges from close to 100 % for $A$ = 175 over more than 90 % for $A$ = 75 to about 20 % for $A$ = 18. The angular transmission for fission fragments is lower. They can only pass the separator when they are emitted close to the forward or backward direction.

In figure 1, the most neutron-rich isotopes predominantly appear in two narrow velocity ranges, corresponding to fission in forward and backward direction, respectively. The most neutron-deficient isotopes, which are produced by fragmentation, however, are found close to the beam velocity. The separation between fragmentation and fission residues is performed in the present experiments by the pattern in velocity and neutron excess [16, 18], as demonstrated in figure 1. The production of the same nuclides by fission and fragmentation allows for directly comparing their velocity distributions, leading to a very interesting result. The fission products result from fission of very heavy nuclei close to the projectile, produced in very peripheral collisions or by electromagnetic excitations, as deduced from an analysis of their velocity distributions [16]. Although the fragmentation products are expected to be produced in more violent collisions, they appear with higher average velocities. This is particularly striking for selenium and zirconium.

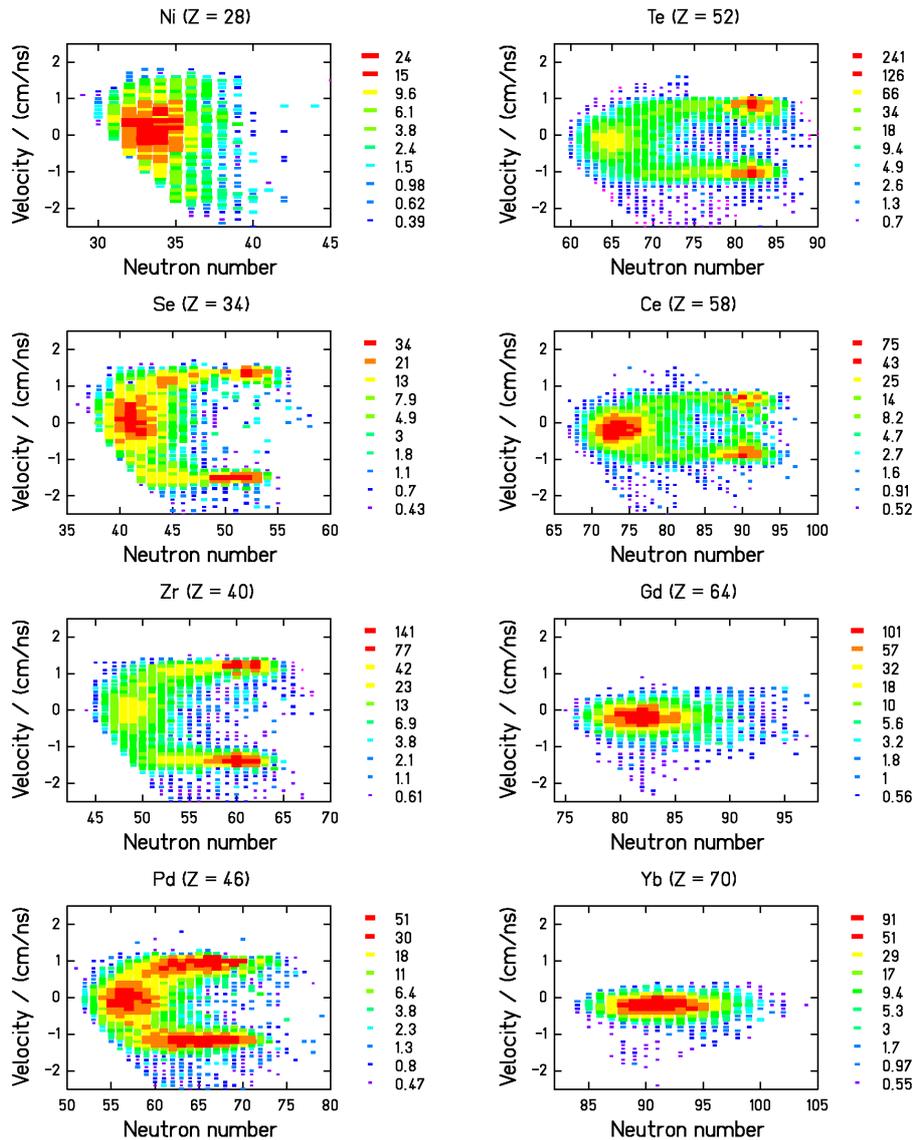

Figure 1: Velocity distributions of the isotopes of eight elements produced in the reaction $^{238}$U + Pb at 1 $A$ GeV. The velocities are given in the projectile frame with the projectile velocity taken to be that of the projectile after traversing half of the target thickness. The measured velocities of the residues behind the target, deduced from the magnetic deflection in the first half of the spectrometer, have been corrected for the energy loss of the fragments in the second half of the target. The scales of the cluster plots are given.

Figure 2 shows the velocity distributions for several isotopes, mostly produced by fragmentation, measured in the reactions $^{238}$U + Pb and $^{238}$U + Ti. The velocity distributions for Z below 28 in the reaction $^{238}$U + Ti were incompletely covered for single isotopes. Since the velocity distributions were found to vary smoothly with neutron number and atomic number (see figure 1), these distributions have been completed from the measured velocity distributions of neighbouring isotopes and elements after weighting them with their corresponding measured cross sections.

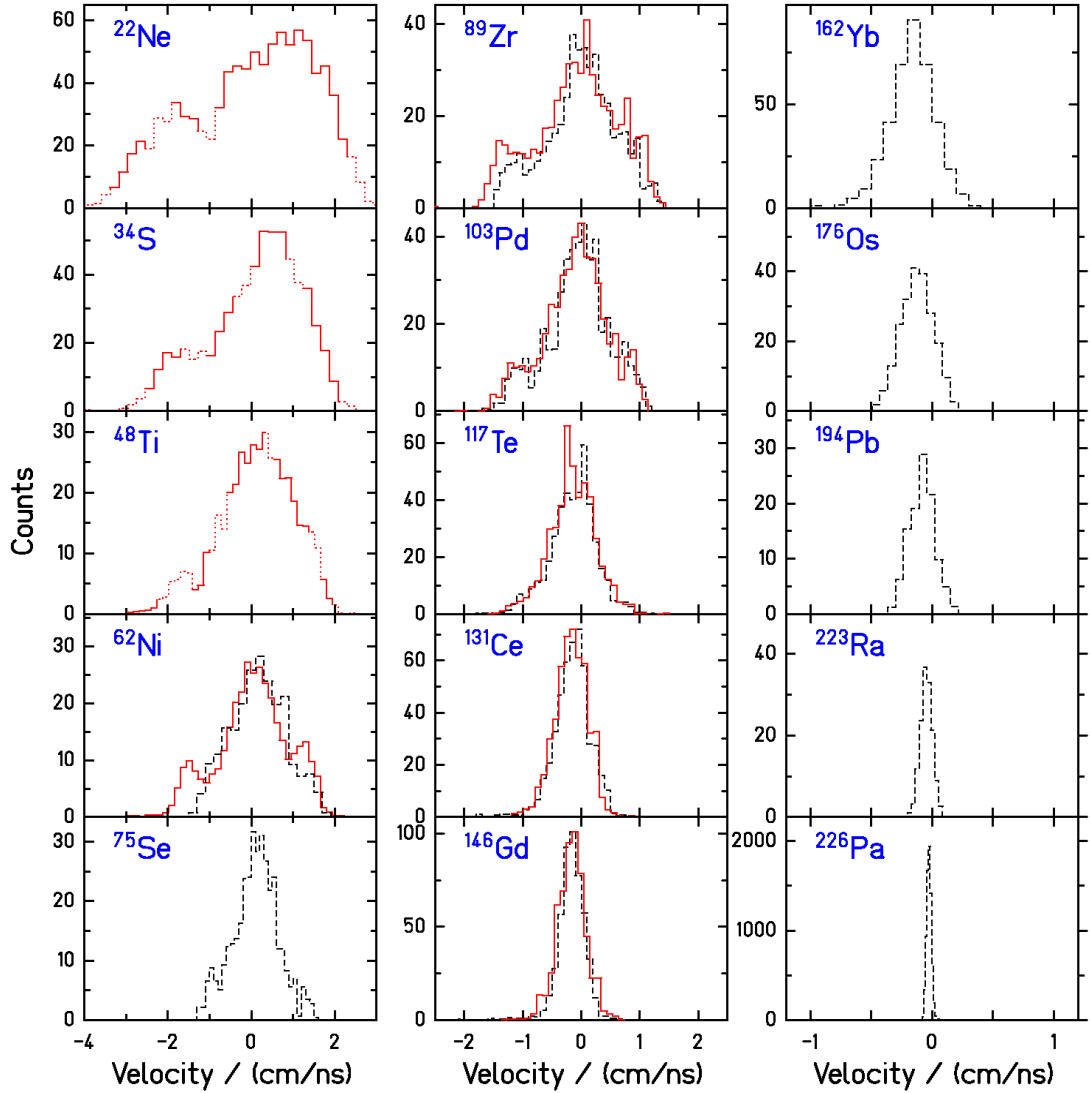

Figure 2: Velocity distributions of different nuclides produced in $^{238}$U + Pb (dashed lines) and $^{238}$U + Ti (full lines) at 1 $A$ GeV measured with the fragment separator [12]. The velocities are given in the frame of the projectile. Note the different scales in velocity. The nuclides were selected from the middle of the fragmentation distributions in figure 1. The dotted parts of the velocity distributions of the lightest nuclei in the system $^{238}$U + Ti are deduced from the data of neighbouring isotopes and elements (see text).

The main peak, close to the zero velocity, can be attributed to fragmentation-evaporation reactions. In the range from titanium to palladium, weak shoulders on both sides of the main peak appear. They originate from fragmentation-fission reactions. Note that these beta-stable or slightly neutron-deficient isotopes are only weakly produced by fission, see figure 1. The distributions of sulphur and neon show two components, one with strongly positive and one with strongly negative velocity. The second one becomes more important for the lightest elements. Since the mean velocity of this component follows the trend of the backward fission peak in the heavier elements, it cannot be excluded that this component results from fission-like binary decay [19] with the corresponding forward component contributing to the main peak. Therefore, we only deduce the mean velocity values of fragmentation-evaporation residues for nuclei with $Z \geq 20$, where the peak of forward fission is clearly separated in velocity from the fragmentation peak. Above $Z = 70$ one Gaussian curve, and below three Gaussian curves have been fitted to the velocity distributions of individual nuclides. The mean velocity values of the fragmentation products were determined from the positions of the corresponding Gaussians. The values shown in figure 3 are averaged over several isotopes, weighted with their production cross sections, traced as a function of the fragment mass. As can be seen in figure 1, the velocity distributions of fragmentation vary only very little over the isotopes of one element. The same is true for the different isobars for a given mass. The fluctuations of the mean velocities of individual isotopes around a smooth trend as well as the fluctuations of the average values shown in figure 3 are within the statistical uncertainties of the fit result. The values are corrected for the enhancement of the transmission for fragments with higher velocities. This correction turns out to be very small, it amounts to 0.0003 cm/ns for $A = 162$, to 0.015 cm/ns for $A = 61$, and to 0.035 cm/ns for $A = 40$. For the system $^{238}$U + Pb, the mean-velocity values of the fragments with A>175 are established by associating the measured mean velocities of the fissioning nuclei to its mass, deduced from the measured velocities of the fission fragments, assuming a liquid-drop description of the total kinetic energy [16]. The absolute uncertainty amounts to less than 0.05 cm/ns for each system, $^{238}$U + Pb and $^{238}$U + Ti, independently.

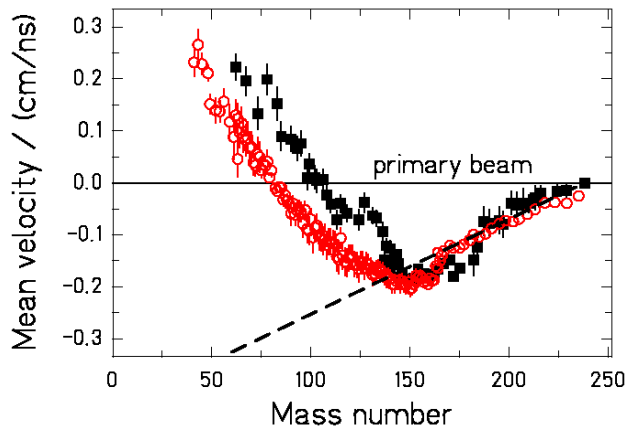

Figure 3: Mean values of the velocity distributions of reaction residues, excluding fission, produced in $^{238}$U + Pb [16] (full squares) and $^{238}$U + Ti (open points, this work) at 1 *A* GeV in the frame of the projectile. Relative uncertainties are shown. The absolute uncertainty amounts to less than 0.05 cm/ns for each system, $^{238}$U + Pb and $^{238}$U + Ti, independently. The dashed line marks the Morrissey systematics [8].

*IV. Discussion.* -- The high-precision measurement of the velocity distributions of the remnants of the projectile in the reactions $^{238}$U + Pb and $^{238}$U + Ti at 1 *A* GeV revealed a surprising result: The velocities of the fragmentation products do not decrease any more if the mass loss becomes large. The velocities of the very light fragments even tend to increase, until finally they are even faster than the projectiles. This finding sheds a new light on the systematics of previously measured data, in particular on the velocities of residues far from the projectile, which did not show a general scaling as a function of mass loss: According to ref. [7], the velocity of the spectator is modified by the expansion of the fireball, and, therefore, other parameters like the mass of the target nucleus and the beam energy have a strong influence. This might explain the large spreading in the Morrissey systematics of previous data for light fragments, including those discussed in the introduction. Our experiments confirm the re-acceleration of the spectator fragment by the participant blast, postulated by Shi, Danielewicz and Lacey [7]. In the framework of this theoretical work, the precise measurement of the kinematical properties of the spectators represents a new tool to determine the in-medium nucleon-nucleon interactions. The present experiment supports the feasibility of this method. As stated in ref. [7], this kind of data will give new information on the equation of state of nuclear matter.

In particular, the acceleration found for relatively small impact parameters, leading to the production of light reaction residues [20], is sensitive to the momentum-dependence of the nuclear mean field [7]. The effects observed in the present experiment can only be compared to the available calculations of ref. [7] for the systems $^{124}$Sn + $^{124}$Sn from 250 to 1000 $A$ MeV and for $^{197}$Au + $^{197}$Au at 1000 $A$ MeV. These calculations predict a sizeable post-acceleration effect. The post-acceleration effect in the calculations amounts to 10 to 20 $A$ MeV/c in the center-of-mass system, corresponding to about 0.25 to 0.5 cm/ns in velocity in the projectile frame, if a momentum-dependent mean field is used, compared to calculations with a momentum-independent mean field. This is in the order of magnitude of the positive velocity values observed in the present experiment for the lightest fragments.

The predicted acceleration effect is rather sensitive to the size of the system. Therefore, one might be surprised about the relatively small difference in the measured acceleration effects of the two systems. However, for a given size of the projectile spectator, the abrasion model [21] predicts that the impact parameter of the system $^{238}$U + Ti is appreciably smaller than the one of the system $^{238}$U + Pb, while the size of the fireball only differs by about 20%. For a more quantitative assessment, dedicated calculations of the systems studied in the present work and in previous experiments performed with different techniques and instruments would be required. According to the model calculations of Shi, Danielewicz and Lacey [7], the peculiar nature of the longitudinal momentum as an observable is the selective sensitivity to the momentum dependence of the mean field. This property is rather unique compared to most experimental signatures, which are sensitive to both, the hardness of the equation of state and the momentum dependence of the mean field.

*V. Conclusion.* -- Applying a high-resolution spectrometer to precisely analyse the velocity distributions of projectile fragments from relativistic heavy-ion collisions revealed a systematic deviation from the expected trend. The mean velocities of the lighter fragmentation products tend to level off and finally increase again with increasing mass loss. The lightest fragments are found to be even faster than the projectiles. This finding is in contrast to the expectation that lighter products, which are assumed to be produced in more violent collisions, are slowed down in the reactions, since they experience more friction in the nucleus-nucleus collision. It appears that the postulated response of the spectators to the participant blast has been established experimentally. Following the ideas presented in a recent theoretical work, the apparent re-acceleration of the projectile spectator will provide a new tool to investigate the equation of state of nuclear matter. According to these calculations, the longitudinal momentum is selectively sensitive to the momentum dependence of the nuclear force.

Discussions with P. Danielewicz are gratefully acknowledged. The work has been supported by the European Community programme "Access to Research Infrastructure Action of the Improving Human Potential" under the contract HPRI-1999-CT-00001.